# Luminous Quasars in Luminous Early-Type Host Galaxies

K. K. McLeod

Smithsonian Astrophysical Observatory, 60 Garden St. MS12, Cambridge, MA 02138

and

G. H. Rieke

Steward Observatory, University of Arizona, Tucson, AZ 85721

## ABSTRACT

Some recent observations of nearby quasars with HST have apparently failed to detect host galaxies. We review the HST observations as well as near infrared ground-based observations of the same objects. We find that the quasar hosts can be detected in the HST data if they are smoothed sufficiently to reveal low surface brightness. The smooth hosts are very difficult to detect with HST but are more easily visible in the deeper, ground-based IR images. The $V-H$ colors obtained by combining the HST and IR data are compatible with normal galaxy colors at the redshifts of the quasars. This behavior can be explained if the hosts are massive early-type galaxies. All together, HST images have been reported for 15 luminous quasars, approximately 13 of which have smooth early-type hosts. This kind of galaxy therefore appears to be the most common host for a luminous quasar.

*Subject headings:* quasars: general–galaxies: photometry–infrared: galaxies

## 1. Introduction

Since Kristian's (1973) pioneering study of the extended emission around quasars, many astronomers have used visible imaging to investigate the "fuzz" properties. By restricting samples to very low redshifts ($z \lesssim 0.5$) and by using linear electronic detectors, studies in the last decade have made significant progress (e.g. Gehren et al. 1984; Malkan 1984; Malkan, Margon, & Chanan 1984; Smith et al. 1986; Hutchings 1987; Hutchings, Janson, & Neff 1989; Romanishin & Hintzen 1989; Véron-Cetty & Woltjer 1990). These studies have determined that quasar fuzz generally has the size, luminosity, and visible color of a galaxy at the redshift of the nucleus and has therefore become known as the "quasar host galaxy." Boroson and collaborators (Boroson, Oke, & Green 1982; Boroson & Oke 1984; Boroson, Persson, & Oke 1985) obtained fuzz spectra more than a decade ago. They argued that the continuum has a contribution from starlight and find probable Mg Ib absorption in several cases. High resolution ($0\rlap{.}''5$) CCD images of the host galaxies have been obtained by Hutchings & Neff (1992) using the rapid guiding HRCam. New large-format IR arrays have also been used to image large numbers of quasar hosts (McLeod & Rieke 1994a,b; Dunlop et al. 1993).

"One of the principal scientific goals for which HST was designed" was to observe quasar host galaxies at unprecedented resolution (Bahcall, Kirhakos, & Schneider 1994a). Three groups have described imaging studies of low-redshift quasars with the optically corrected WFPC2 aboard HST: Bahcall et al. (1994a, 1995a,b,c); Hutchings and co-workers (Hutchings et al. 1994, Hutchings & Morris 1995); and Disney et al. (1995). All together, results are reported for 18 quasar/host galaxy systems, 15 of which have luminous nuclei. A large range of host galaxy properties is represented, ranging from highly disturbed and/or interacting morphologies through galaxies with large scale spiral arms to five cases where the host may not be detected.

The incidence of faint or non-detectable hosts for some of the luminous quasars contrasts with expectations from groundbased work that virtually all luminous quasars lie in luminous hosts. In addition to the possibility of observational or analytical problems with the various data sets, three hypotheses have been suggested to account for the HST results: 1.) some luminous quasars lie in very low mass galaxies, or perhaps in no galaxies at all; 2.) some quasar hosts are significantly redder in V - H than normal galaxies; or 3.) some hosts have very smooth morphologies that offer no structures easily detectable by HST. Although indirect arguments have been offered in favor of the third possibility (Hutchings 1995), the evidence will remain circumstantial until both groundbased and HST data on the identical set of quasars can be combined to yield a consistent explanation.

In this paper, we discuss our previous and new groundbased data and re-analyze the HST data of objects for which we have near-IR data to test what properties of the data sets and/or the host galaxies can account for the HST non-detections.

## 2. Infrared Data and Analysis

### 2.1. Previous Data

We begin by reviewing the results of our near-IR study of low-redshift quasars. We note that our results are perfectly consistent, albeit within considerable uncertainties, with the ground-based CCD studies mentioned above (for details see McLeod & Rieke 1994a,b, 1995). We chose to observe our sample of quasars in the H band (1.6$\mu$m) using a 256x256 NICMOS array camera on the Steward 90" telescope. Figure 1 shows why this wavelength is in an ideal regime for detecting the host galaxy—the output from galaxy starlight peaks exactly where most quasars have a local minimum in their energy distributions. Thus, for quasars at redshift $z \approx 0.2$, H band gives much better contrast of starlight to nuclear light than does the V band.

We obtained images at a resolution of $\lesssim 1\rlap.{''}5$ of the 50 quasars from the Palomar Green survey

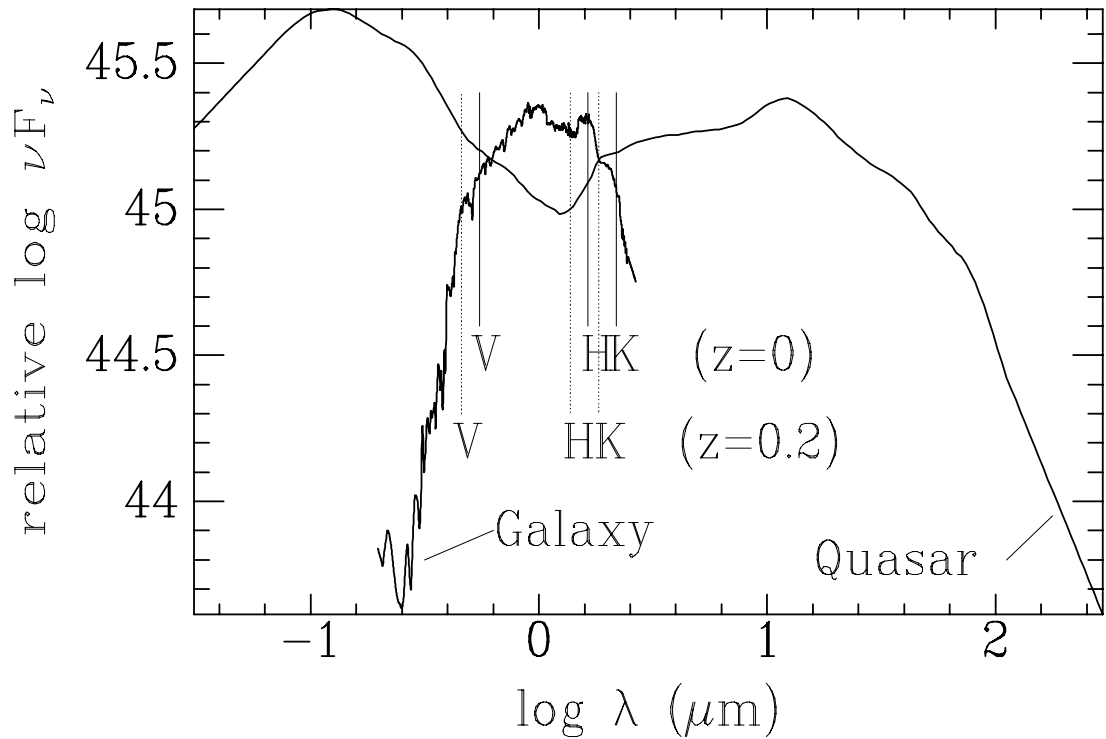

Fig. 1.— Spectral energy distributions of typical galaxies (M. Rieke, private communication) and quasars (Elvis et al. 1994). The normalization is arbitrary; for a luminous quasar, the nucleus is much brighter than the galaxy. H-band observations of quasars allows us to view the host galaxies where their starlight is peaking, and with less contamination from nuclear light than is possible at visible wavelengths.

(PG survey, Schmidt & Green 1983) with $z < 0.3$. We combined many short exposures (less than one minute) in the infrared images, so the nucleus was never saturated. To remove the nuclear contribution, we generated one-dimensional (1-D) radial luminosity profiles of the quasars and of stars from the same frames. The profiles reached the $1\sigma$ sky limit at H$\approx 23$ mag arcsec$^{-2}$, which corresponds to V$\approx 26$ mag arcsec$^{-2}$ for typical galaxy colors. We then scaled the stellar profiles, or PSFs, to match the quasar profiles at the innermost pixel. Next, we subtracted the fraction of scaled PSF needed to make the resulting (quasar − PSF) profile start to turn over in the center. For reasons discussed in our papers, we believe that the remaining light gives a reasonable estimate of the contribution from the host galaxy. As a convenient way of smoothing and adding up this light, we fit the residuals with an exponential profile. Note that we do not claim that the true profiles are best fit by an exponential disk; for high-luminosity quasars, we could not reliably determine whether the profiles were fit better by a disk law, an $r^{1/4}$ law, or neither. If the profiles are actually closer to $r^{1/4}$ laws, our subtraction and fitting procedure leads to an underestimate in the amount of host light.

The quasar was seen to be extended in 45 cases. The resulting host galaxy magnitudes for these quasars, as well as for 50 Seyferts from our larger sample and additional objects from the literature, are shown in Figure 2 (in our previous study and throughout this paper, we take $H_0 = 80$ km s$^{-1}$ Mpc$^{-1}$). The uncertainties in host galaxy magnitudes for low-luminosity AGN are small, but for the most luminous quasars the uncertainties can be up to 0.5 mag. Of the 15 low-redshift, luminous quasars imaged so far by HST, the eight quasars in Bahcall et al. (1995a) plus PG0052+251 from Bahcall et al. (1995c) are included in our sample. All were found to be extended in the near-IR images.

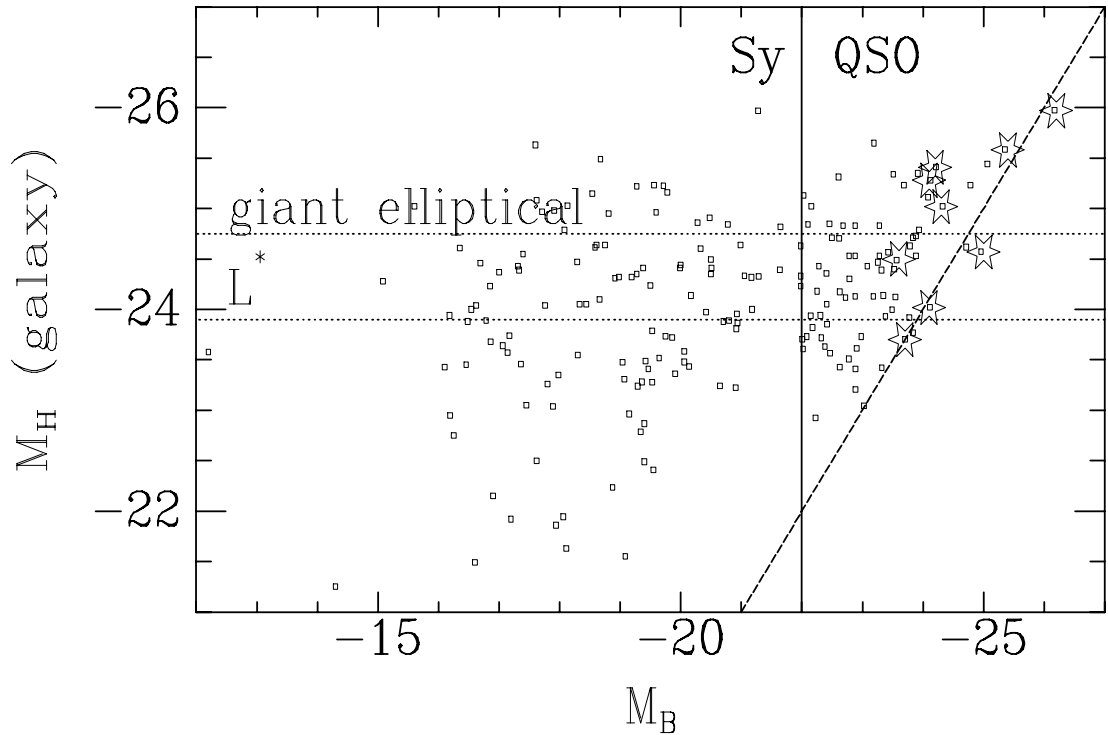

Fig. 2.— Host galaxy v. nuclear absolute magnitudes from McLeod & Rieke 1995 (see that paper for more details; includes points from Dunlop et al. 1993, Kotilainen & Ward 1994, Zitelli et al. 1993, and Granato et al. 1993). The nine luminous HST quasars included in our sample are highlighted by stars.

## 2.2. New Data

Not surprisingly, some of the quasars for which HST did not detect hosts were also among the weaker detections in our data; in fact, we had already noted that PG0953+414 was only marginally resolved. Therefore, we repeated the measurements of these five quasars with the Steward 90" telescope on 1995 March 17-19 (U.T.). Our approach was similar to that used previously, but we took extra care to be sure a suitable star was present on the quasar frame to determine the instrumental PSF. We also integrated on each quasar for about twice as long as previously and, fortunately, the seeing was better, with a nearly pixel-limited PSF FWHM of $\sim 1\rlap.''2$. Our analysis of these data confirms the detections of host galaxies at brightnesses similar to those measured previously in all cases, including for PG0953+414.

Figure 1 suggests another test of the reality of these detections: the expected spectral energy distributions of the quasars and host galaxies have very different slopes between H(1.6$\mu$m) and K(2.2$\mu$m), so the $H - K$ colors of the "fuzz" should be much bluer than for the nucleus. To investigate this possibility, we obtained K-band images for several objects. The cleanest test came from the H and K images of PG1307+085, which were obtained under identical seeing. For this object the fuzz color was $H-K \simeq 0.56$, much closer to the color expected for a normal galaxy at this redshift ($H-K$=0.43) than to the measured nuclear color ($H-K$=1.04). For the other objects, the results are qualitatively similar.

### 2.3. Host Galaxy Masses

As seen in Figure 1, near-IR light gives a better measure than blue or visible light of the bolometric luminosity and hence of the total stellar mass of the galaxy, and its interpretation is relatively immune to the effects of interstellar extinction. Figure 2 shows that Seyferts and lower luminosity quasars are often found in galaxies with H luminosities much less than that of an $L^*$ galaxy (i.e. a galaxy at the knee in the Schechter function description of the local field galaxy luminosity function; our value is derived from Mobasher, Sharples, & Ellis 1993). For "luminous quasars," which we take here to satisfy $M_B(nucleus) \lesssim -23$, there appears to be a linear relation between $M_B$(nucleus) and the minimum host galaxy magnitude $M_H$; this is indicated by the diagonal line in Figure 2. Quasars with $M_B(nucleus) \lesssim -24$ appear not to occur in sub-$L^*$ hosts.

The nine HST quasars for which we have infrared images are denoted by stars on Figure 2. It is noteworthy that four of the five of the galaxies listed by Bahcall et al. (1995a) as non-detections lie within 0.5 mag of the diagonal line. The large difference in brightness between the faint extended galaxy and the bright nucleus undoubtedly contributes to the difficulty in detecting these galaxies with HST. Nonetheless, their host galaxies in general do not stand out as being underluminous in $M_H$ relative to other quasar hosts, nor relative to luminous galaxies without AGNs. These quasars lie in galaxies of near-IR luminosity, and hence mass, typical both of other quasar host galaxies and of luminous galaxies in the field.

## 3. Re-Analysis of HST Images

The obvious question is, "If the galaxies are there and are as bright as the infrared images imply, why were the Bahcall et al. upper limits so faint?" The answer to this question reveals new information on the structures of the host galaxies.

### 3.1. Detection of Host Galaxies

For three objects, 3C273, PG1116, and PG1444, Bahcall et al. (1995a) detect extended emission that is plausibly interpreted as the host galaxy. We have made additional tests that support this interpretation. Comparing our IR images to the PSF-subtracted HST images, we find that, although the HST frames do not reach as faint a surface brightness as the IR images, the faint fuzz shows the same shapes in both sets of data. A good example is PG1444+407, which shows an elongation of the outer isophotes in the north-south direction in both images. Based on our luminosity profile and assuming normal galaxy colors, and considering the quoted sky limit of the HST observations ($V \approx 24.7$ mag arcsec$^{-2}$) we would predict that the HST fuzz should be visible to approximately $4\rlap{.}''5$ radius, which is indeed where it disappears.

Another test comes from the case of PG1116+215. In Figure 3 we show an image of this quasar obtained from the HST archive (upper left). We have subtracted the PSF from PG1116 according to guidelines outlined by Bahcall et al. Although Bahcall et al. suggest the host of this galaxy is at or below $L^*$, the residual light looks nearly identical to the nearby galaxy, which has luminosity $L \approx 1.4L^*$ if at the redshift of the quasar. Note that we had to stretch the gray scale logarithmically to see the low surface brightness parts of both galaxies.

For the five other galaxies in their paper, Bahcall et al. claim not to detect fuzz. They have done a very careful job of subtracting out the PSF from these images. The results, shown in their Figure 4, indicate faint residuals around the quasar. Bahcall et al. 's tests of stars-minus-stars (see their Figure 10) do not show similar residuals. We conclude that the outer residuals may show the hosts in these cases. Though the HST images do not go as deep as our IR images, we are able to see shape agreement even in these difficult cases. In PG0953, for example, we find in both of our images (Figure 1 in McLeod & Rieke 1994b and our new image) an asymmetrical extension to the southwest, in agreement with the Bahcall et al. Figure 4. In addition, we have analyzed the archive image of PG1202 and find, similar to the case we show in Figure 3, an underlying structure nearly identical to that of its own close galaxy companion.

### 3.2. Systematic Errors and Host Galaxy Colors

The fits of a disk+PSF to the quasar images presented in Bahcall et al. allow a quantitative comparison because this analysis technique is close to the procedure used with the IR data. A comparison of results is shown in Table 1. The "normal" $V-H$ colors were computed assuming $V-H \sim 3$, appropriate for all but the latest-type galaxies (Griersmith, Hyland, & Jones 1982), plus a k-correction assuming the galaxies have a visible energy distribution similar to that shown in Figure 1. Listed in the table are the Bahcall et al. F606W magnitudes, which are not exactly the same as V magnitudes; depending on the intrinsic visible colors of the host, the filter transformations listed in Bahcall et al. (1994b) indicate that the V magnitude can be up to 0.4 mag fainter than the F606W magnitude. The galaxies appear to be slightly redder than normal galaxies.

Table 1: HST v. IR host galaxy magnitudes

| Object | $m_{F606W}$[a] | H[b] | $m_{F606W} - H$ measured | V–H normal |
|---|---|---|---|---|
| PG0052+251 | 16.8 | 14.5 | 2.3 | 3.16 |
| PG0953+414 | 19.5 | 15.4 | 4.1 | 3.32 |
| PG1116+215 | 18.1 | 14.0 | 4.1 | 3.20 |
| PG1202+281 | 18.7 | 15.1[c] | 3.6 | 3.18 |
| PG1226+023 (3C273) | 16.8 | 13.0 | 3.8 | 3.17 |
| PG1302−102 (PKS) | 18.4 | 14.8[c] | 3.6 | 3.47 |
| PG1307+085 | 19.0 | 15.2 | 3.8 | 3.16 |
| PG1444+407 | 18.8 | 15.2 | 3.6 | 3.42 |
| PG1545+210 (3C323.1) | 19.1 | 14.8[c] | 4.3 | 3.42 |

[a]Bahcall et al. 1995a,c exponential disk fits; depending on the host's intrinsic color, the V magnitude could be fainter than $m_{F606W}$ by up to 0.4 mag
[b]McLeod & Rieke 1994b
[c]Includes some light from close companions

John Bahcall has generously provided us with the radial profiles of PG1116+215 and of the PSF star. We put these profiles through the same procedure we used with the IR data, differing in one detail only: because the HST profiles are saturated in the center, we normalized them just outside the saturation region. The resulting galaxy magnitude was $m_{F606W} \sim 17.8$, about 0.3 mag brighter than the Bahcall et al. 2-D analysis determination, and the profile had the same e-folding length as did the IR profile. When we extracted our own profiles from the archived images, we found we could reach a sky brightness of $V \approx 25.5$ mag arcsec$^{-2}$. Putting our profiles through the same reduction, the inferred galaxy magnitude was brighter still, $m_{F606W} \sim 17.5$. Of course it would be preferable when estimating the fuzz luminosity in the HST images to just add up the residual light from the PSF-subtracted image directly. This is unfortunately very difficult because of the complicated PSF and the many artifacts left after PSF subtraction. We have attempted to mask out all of these features and add up the light remaining after PSF subtraction for the archived image of PG1116+215. The resulting host brightness, which we regard as very uncertain, is $m_{F606W} \sim 17.5$, consistent with our radial profile analyses above.

That is, using procedures as close as possible to the reductions used on our infrared images, i.e. pushing to lower surface brightness levels by taking azimuthal averages, we infer host galaxy magnitudes 0.3 to 0.6 mag brighter than the brightest value assigned by Bahcall et al. This result emphasizes the difficulties in measuring accurate host galaxy magnitudes; systematic errors approach 0.5 mag between "equivalent" reduction techniques applied to the identical data. Similar systematic errors are likely to occur in the infrared reductions, although we do not have the ability to apply the procedures of Bahcall et al. to them. Another important source of bias is the presence of close companion galaxies (such as for PG1202), which would be merged with the host in the infrared but would usually be separated from the fits to the HST data.

Both sources of bias we have identified would tend to make a comparison of HST-determined V and groundbased H tend toward the red, with V - H perhaps 0.5 mag or more redder than the true galaxy colors. We previously concluded from a comparison with groundbased visible data that the host galaxies might have normal to bluer than normal colors, albeit with large uncertainties (McLeod & Rieke 1994b). The contrary indication from the HST data emphasizes the uncertainties. We conclude that the quasar host galaxies have visible-to-infrared colors that are, within the errors, the same as for normal galaxies.

### 3.3. Host Galaxy Morphology

We have shown that the Bahcall et al. work gives results compatible with the ground-based results if analyzed in a way that discards all the gain in resolution afforded by HST. However, there is still the question of why Bahcall et al.'s sensitivity tests gave upper limits on the luminosities that are lower than determined above. For their tests, they simulated quasar+host galaxy images by putting a point source on top of images of eight different kinds of galaxies found in the WFPC2 frames (see their Figure 5). They then progressively decreased the galaxy fluxes until they could no longer detect the galaxies under the point source. Most of their test galaxies have sharp, high surface brightness features that make detection easier and hence make the limits more stringent. However, as Bahcall et al. point out, the detection limits based on fairly smooth galaxies are not as faint. The smooth test galaxies (c) – an S0 – and (d) – an elliptical – from that figure give limits closer to the luminosities derived from the disk fits and from the IR data. Note that galaxy (c) is the same galaxy shown as a comparison for PG1116 in our Figure 3. Thus, we believe these tests have shown us that most of these quasars are found in *smooth*, round, and early-type galaxies.

### 4. Conclusions

We conclude that the high-luminosity quasars imaged by Bahcall et al. (1995a) lie in host galaxies with normal visible-to-infrared colors. The hosts span a range of luminosities but are generally at or above the $L^*$ level. Note, however, that the galaxy luminosities derived from

# – 9 –

both HST and ground-based near-IR images are uncertain due to the relatively broad point spread function in the ground-basd images, the relative shallowness of the HST images, and the uncertainties of PSF subtraction in both.

The sensitivity tests carried out by Bahcall et al. (1995a) for differing host galaxy types indicate that the high luminosities of these galaxies can be reconciled with the difficulties in detecting them with HST if they are smooth and round early-type galaxies. The host morphologies seen in HST images of the high-luminosity quasars from the Disney et al. (1995), Hutchings et al. (1994), and Hutchings and Morris (1995) also are indicative of early-type galaxies. Only a few (e.g. PKS 2349−014, Bahcall et al. 1995b; and PG 0052+251, Bahcall et al. 1995c) of the 15 luminous quasars imaged by HST appear to lie in galaxies with high surface brightness structure that may indicate later types. Although better statistics on this tendency will be accumulated as more data are acquired with HST, it appears that the data already in hand show a strong preference for high luminosity quasars to lie in early-type galaxies.

We thank John Bahcall for interesting discussions, for offering to share ideas as well as data, and especially for pointing HST at these quasars. We thank the referee for promptly delivered comments that improved the readability of this paper. We also thank John Hutchings and Richard Elston for helpful conversations. We thank Brian McLeod for help with the HST Archive data, and we thank Marcia Rieke and Earl Montgomery for maintaining the excellent IR camera. KKM gratefully acknowledges support from B. J. Wilkes through NASA grant NAG W-3134. This work was also supported under NSF grant AST91-16442 to GHR.

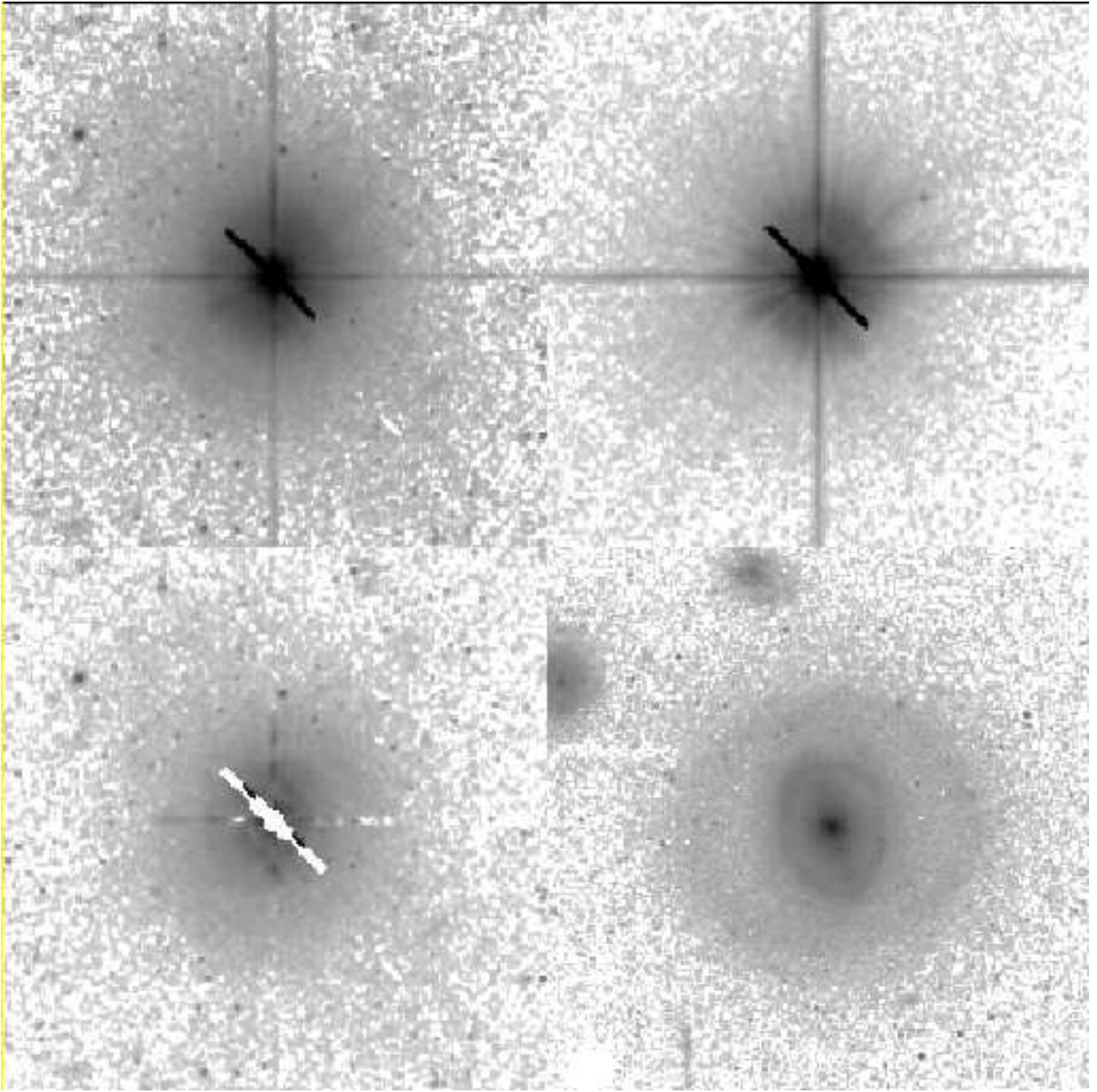